\documentclass[10pt,letterpaper]{article}
\usepackage[numbers]{natbib}

\usepackage{amsmath} 
\usepackage[version=4]{mhchem}	
\usepackage{xcolor}

\usepackage{graphicx}
\usepackage[export]{adjustbox}
\usepackage[utf8]{inputenc}
\usepackage{nameref,hyperref}
\usepackage[left]{lineno}
\usepackage{microtype}
\DisableLigatures[f]{encoding = *, family = * }

\usepackage{setspace} 

\usepackage{changepage}

\usepackage[aboveskip=1pt,labelfont=bf,labelsep=period,singlelinecheck=off]{caption}

\makeatletter
\renewcommand{\@biblabel}[1]{\quad#1.}
\makeatother

\usepackage{lastpage,fancyhdr,graphicx}
\usepackage{epstopdf}
\usepackage{color}
\definecolor{Gray}{gray}{.25}
\usepackage{graphicx}
\usepackage{sidecap}
\usepackage{wrapfig}
\usepackage[pscoord]{eso-pic}
\usepackage[fulladjust]{marginnote}
\reversemarginpar
\definecolor{red}{cmyk}{0,0,0,1}
\begin{document}
	\vspace*{0.35in}
	\begin{flushleft}
		{\Large
			\textbf\newline{Circular spectropolarimetric sensing of higher plant and algal chloroplast structural variations}
		}
		\newline
		\\
		C.H. Lucas Patty\textsuperscript{1*},
		Freek Ariese\textsuperscript{2},
		Wybren Jan Buma\textsuperscript{3},
		Inge Loes ten Kate\textsuperscript{4},
		Rob J.M. van Spanning\textsuperscript{5},
		Frans Snik\textsuperscript{6}
		\\
		\bigskip
		\tiny 1 Molecular Cell Physiology, VU Amsterdam, De Boelelaan 1108, 1081 HZ Amsterdam, The Netherlands
		\\
		{2} LaserLaB, VU Amsterdam, De Boelelaan 1083, 1081 HV Amsterdam, The Netherlands
		\\
		{3} HIMS, Photonics group, University of Amsterdam, Science Park 904, 1098 XH Amsterdam, The Netherlands	
		\\
		{4} Department of Earth Sciences, Utrecht University, Budapestlaan 4, 3584 CD Utrecht, The Netherlands
		\\
		{5} Systems Bioinformatics, VU Amsterdam, De Boelelaan 1108, 1081 HZ Amsterdam, The Netherlands
		\\
		{6} Leiden Observatory, Leiden University, P.O. Box 9513, 2300 RA Leiden, The Netherlands
		\\
	
		\bigskip
		
		\bf *lucas.patty@vu.nl
		
	\end{flushleft}
	 
	\section*{Abstract}
	Photosynthetic eukaryotes show a remarkable variability in photosynthesis, including large differences in light harvesting proteins and pigment composition. \textit{In vivo} circular spectropolarimetry enables us to probe the molecular architecture of photosynthesis in a non-invasive and non-destructive way and, as such, can offer a wealth of physiological and structural information. In the present study we have measured the circular polarizance of several multicellular green, red and brown algae and higher plants, which show large variations in circular spectropolarimetric signals with differences in both spectral shape and magnitude. Many of the algae display spectral characteristics not previously reported, indicating a larger variation in molecular organization than previously assumed. As the strengths of these signals vary by three orders of magnitude, these results also have important implications in terms of detectability for the use of circular polarization as a signature of life.
	
	\section*{Keywords}
	Circular polarization, photosynthesis, chloroplast, chlorophyll, algae
	\begin{center}
		
	\end{center}
	
	\nolinenumbers\clearpage
\section{Introduction}
Terrestrial biochemistry is based upon chiral molecules. In their most simple form these molecules can occur in a left-handed and a right-handed version called enantiomers. Unlike abiotic systems, nature almost exclusively uses these molecules in only one configuration. Amino acids, for instance, primarily occur in the left-handed configuration while most sugars occur in the right-handed configuration. This exclusive use of one set of chiral molecules over the other, called homochirality, therefore serves as a unique and unambiguous biosignature \cite{Schwieterman2017}. 

Many larger, more complex biomolecules and biomolecular architectures are chiral too and the structure and functioning of biological systems is largely determined by their chiral constituents. Homochirality is required for processes ranging from self-replication to enzymatic functioning and is therefore also deeply interwoven with the origins of life.

The phenomenon of chirality, i.e. the molecular dissymmetry of chiral molecules, causes a specific response to light \cite{Fasman2013, Patty2018a}. This response is both dependent on the intrinsic chirality of the molecular building blocks and on the chirality of the supramolecular architecture. Polarization spectroscopy enables these molecular properties to be probed non-invasively from afar and is therefore of great value for astrobiology and the search for life outside our solar system. Polarization spectroscopy also has a long history in biological and chemical sciences. Circular dichroism (CD) spectroscopy utilizes the differential electronic absorption response of chiral molecules to left- and, right-handed circularly polarized incident light and is very informative for structural and conformational molecular dynamics. As such it has proven to be an indispensable tool in (bio-)molecular research. 

Chirality can also be observed in chlorophylls and bacteriochlorophylls utilized in photosynthesis. While their intrinsic CD signal is very weak due to their almost planar symmetrical structure, these chlorophylls are organized in a chiral supramolecular structure that greatly enhances these signals \cite{Garab2009}. This is particularly the case for the photosynthetic machinery in certain eukaryotes, where photosynthesis is carried out in specialized organelles, chloroplasts, which in higher plants have a large molecular density yielding anomalously large signals: polymer and salt induced (psi)‐type circular dichroism \cite{Keller1986, Garab2009, Garab1991, TinocoJr1987}.

While circular dichroism spectroscopy depends on the modulation of incident light to detect the differential extinction of circularly polarized light, we have recently shown that in leaves comparable results can be obtained by measuring the induced fractional circular polarization of unpolarized incident light \cite{Patty2017, Patty2018b}. As the latter only requires modulation in front of the detector it offers unique possibilities, allowing to probe the molecular architecture from afar. In vegetation, the influence of photosynthesis functioning and vegetation physiology on the polarizance could provide valuable information in Earth remote sensing applications, as was demonstrated for decaying leaves \cite{Patty2017}. As homochirality is a prerequisite for these signals (left- and right-handed molecules display an exactly opposite signal and will thus cancel out each other if present in equal numbers) and is unique to nature, circular polarization could also indicate the unambiguous presence of life beyond Earth and as such is a potentially very powerful biosignature \cite{Sparks2009, Sparks2009a, Wolstencroft1974, Patty2018a, Pospergelis1969, Schwieterman2017}.

Higher plants evolved relatively recently in contrast to microbial life. Biosignatures of microbial life are mostly focused on in astrobiology (and which also display typical circular polarization signals \cite{Sparks2009}). While molecular analysis suggests higher plants appeared by 700 Ma \cite{Heckman2001}, the earliest fossil records date back to the middle Ordovician ($\sim$ 470 Ma) \cite{Wellman2000}. The earliest microbial fossil records date back to 3.7 Ga \cite{Nutman2016} and oxygenic photosynthesis (in cyanobacteria) is likely to have evolved before 2.95 Ga \cite{Planavsky2014}. It is however unclear if photosynthetic microbial life would be able to colonize terrestrial niches extensively enough to be used as a remotely detectable biosignature. 

\begin{figure}[ht]
	\includegraphics[width=1.1\textwidth]{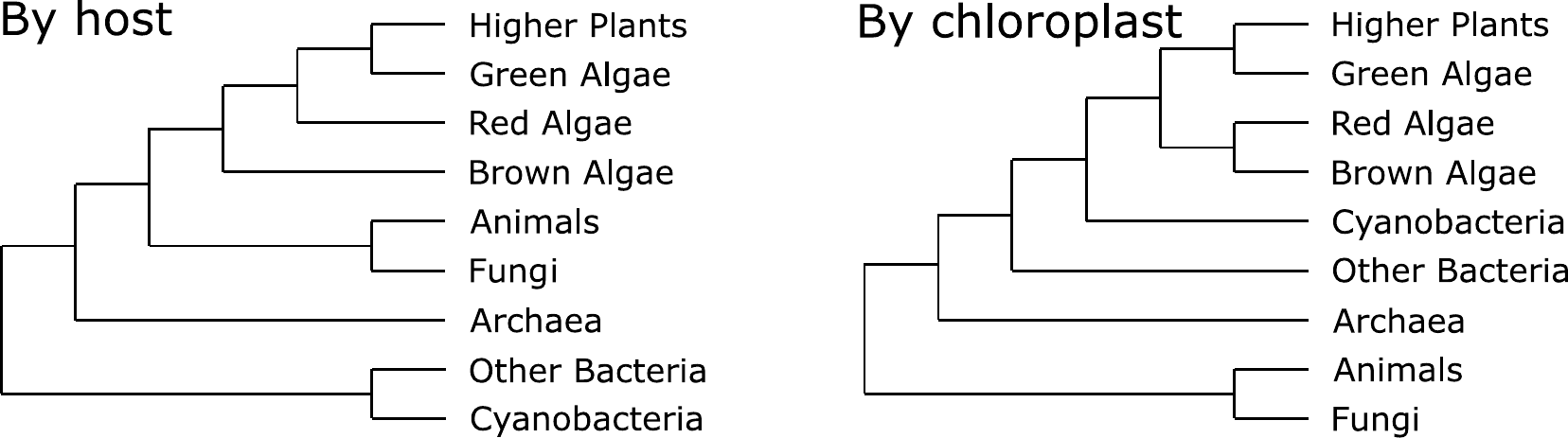}
	\caption{Evolutionary relationships based on the host rRNA (left) and based on chloroplast DNA (cpDNA) (right)}
	\label{fig:Evo}
\end{figure}

On the other hand, these photosynthetic bacteria stood at the basis of the evolution of higher plants as their photosynthetic apparatus evolved from a endosymbiosis between a cyanobacterium and a heterotrophic host cell. It is widely accepted that all chloroplasts stem from a single primary endosymbiotic event \cite{Moreira2000, Ponce-Toledo2017, McFadden2001}. Not all photosynthetic eukaryotes, however, descend from this endosymbiotic host, as certain algae acquired photosynthesis through secondary endosymbiosis of a photosynthetic eukaryote \cite{McFadden2001, Green2011}. The simplified evolutionary relations between the different algae, based on the host and on the chloroplasts, are shown in Figure \ref{fig:Evo}.

Although algae contribute up to 40 \% of the global photosynthesis \cite{Andersen1992}, they have received limited attention in astrobiology so far. While not as ancient as microbial life, algae are considerably older than plants, with fossil evidence of red algae dating back to 1.6 Ga \cite{Bengtson2017}. Additionally, molecular research on algae has mainly focused on a few unicellular algae, rather than multicellular species, and systematic studies on the chiral macro-organization of algal photosynthesis are lacking \cite{Garab2009}. Despite the common origin, millions of years of evolution has caused chloroplasts to show a remarkable diversity and flexibility in terms of structure. 

In higher plants the chloroplasts typically display cylindrical grana stacks of 10–20 membrane layers that have a diameter of 300 nm – 600 nm. The stacks are interconnected by lamellae of several hundred nm in length \cite{Mustardy2003}. Additionally, certain plants can display grana stacks of more than 100 membrane layers \cite{Anderson1973, Steinmann1955} while the bundle sheath cells of certain C4 plants, such as maize, lack stacked grana and only contain unstacked stroma lamellae \cite{Faludi-Daniel1973}. 

In higher plants, the psi-type circular polarizance is largely dependent on the size of the macrodomains formed by the photosystem II light-harvesting complex II supercomplexes (PSII-LHCII). The structure of PSII-LHCII in higher plants is relatively well known and consists of a dimeric PSII core complex ($C_{2}$) and associated trimeric LHCII, subdivided in three types based on their position and association with the core: Loose (L), Moderate (M) and Strong (S). Additionally, three minor antennae occur as monomers (CP24, CP26, CP29) \cite{Boekema1999}. The position of trimer L is still unclear and has so far only been observed in spinach \cite{Boekema1999}. The protein constituents and their typical circular polarization signature have been determined by T\'oth et al (\cite{Toth2016}). Furthermore, the negative band of the psi-type split signal is associated with the stacking of the thylakoid membranes, whereas the positive band is associated with the lateral organization of the chiral domains \cite{Garab1988, Garab1991a, Cseh2000}.

The evolutionary history of grana and their functional advantage has been a matter of debate. It has been proposed that the structural segregation by grana of PSII and PSI prevents excitation transfer between these systems \cite{Albertsson2001, Nevo2012, Trissl1993}. The extended compartmentation brought upon by grana might also aid regulatory pathways such as used in carbon fixation \cite{Anderson1999}. It has been suggested that grana facilitate the regulation of light harvesting and enhance PSII functioning from limiting to saturating light levels, while at the same time protecting it from sustained high irradiance \cite{Anderson1999}. Together with other adaptations, it has been hypothesized that these changes might have ultimately enabled green algae/plants to colonize and dominate various terrestrial niches \cite{Nevo2012}. Others have suggested that it might simply be a lack of competition; red algae for instance have probably experienced several evolutionary bottlenecks, vastly decreasing their genome size and therewith their potential for evolutionary adaptation \cite{Collen2013}. 

Most closely related to higher plants are the green algae, which share a quite recent common ancestor. Similar to higher plants, green algae contain chlorophyll \textit{a} and \textit{b}. The structural composition of their photosynthetic machinery and the associated genes are primarily known from the unicellular green algae \textit{Chlamydomonas}. Despite the high sequence similarity there are significant differences between the supercomplexes of higher plants and green algae. Importantly, green algae lack CP24, resulting in a different organization of the PSII-LHCII supercomplex \cite{Tokutsu2012}. While many green algae display thylakoid stacking, which can be up to 7 membrane layers thick \cite{Remias2005}, true grana in green algae are rare and only occur in the late branching taxa Coleochaetales and Charales \cite{Gunning1999, Larkum2003}.

Red algae also contain thylakoid membranes but these are never stacked. Furthermore, unlike green algae and plants, red algae can contain chlorophyll \textit{d}, a pigment with an absorption band from 700 nm to 730 nm \cite{Larkum2005}. The red algae also contain phycobilisomes that serve as the primary antennae for PSII rather than the chlorophyll binding proteins found in higher plants and other algae. These phycobilisomes are homologous to those in cyanobacteria, but are lacking in plants and other algae  \cite{McFadden2001}.

\begin{figure}[ht]
	\hspace{-1 cm}
	\includegraphics[width=1.3\textwidth]{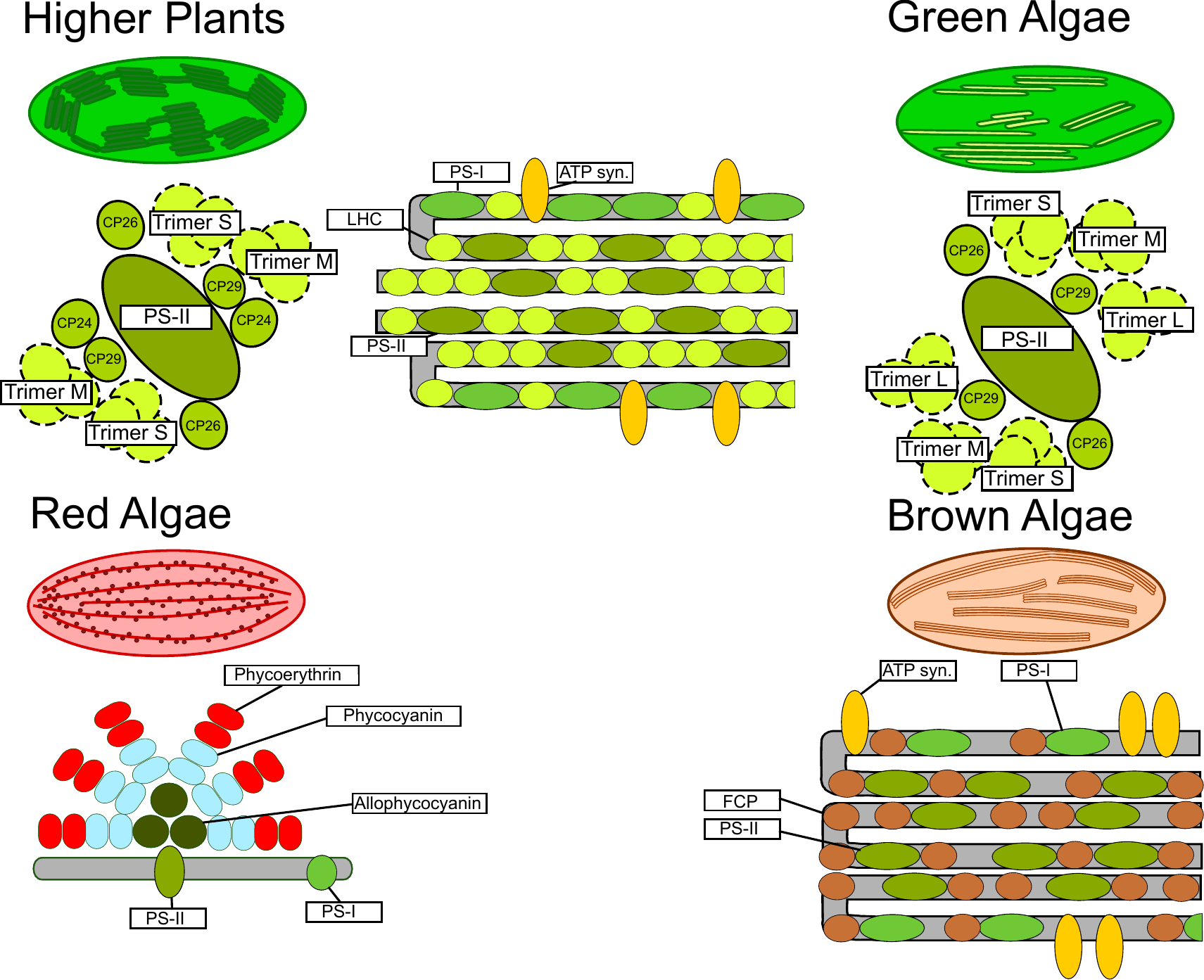}
	\caption{Schematic representation of the photosynthetic structures of higher plants and algae. There is a distinct organizational difference in the supercomplexes between higher plants and algae. Additionally, while green algae display stacked thylakoid membranes, they lack true grana. Red algae contain phycobilisomes, unlike the other algae. In brown algae the thylakoid membranes are threefold and the supercomplex organization is not entirely resolved.}
	\label{fig:CP}
	
\end{figure}
Similarly, brown algae do not possess stacked thylakoid membranes but also do not contain phycobilins. All brown algae contain chlorophyll \textit{a} and usually chlorophyll \textit{c$_1$}, \textit{c$_2$} and/or \textit{c$_3$}. The light-harvesting systems in brown algae are based on fucoxanthin chlorophyll $a/c_{1,2,3}$ proteins (FCP), which are homologous to LHC in higher plants/green algae but have a different pigment composition and organization \cite{Premvardhan2010, Buechel2015}. Although this is still under debate \cite{Burki2016}, the brown algae have been classified as one supergroup \cite{Dorrell2011}. Most brown algae have chloroplasts which were acquired through one or more endosymbiotic events with red algae \cite{Dorrell2011}. Additionally, certain species of brown algae have been shown to display psi-type circular polarizance, although varying magnitudes of these signals have been reported, ranging from very weak to signals similar to higher plants (see \cite{Garab2009} and references therein). 

In the present study we measure the fractional circular polarizance of various {\color{red}higher plants and} multicellular algae. As the level of chiral macro-organization varies greatly between unicellular algae, we expect especially in multicellular algae that the organization can reach a higher or different level of complexity. These studies will additionally assess the feasibility of biosignature detection for (eukaryotic) photosynthesis from different evolutionary stages. While transmission and reflectance generally show a comparable spectral profile, the signals in reflectance are often weaker (e.g. due to surface glint). In the present study we will therefore only display the results in transmission, as it provides better sensitivity for small spectral changes between samples. 

\clearpage

\section{Materials and Methods}

\subsection{Sample collection}
\textit{Ulva lactuca},\textit{ Porphyra sp.} and \textit{Saccharina latissima} were grown in April at the Royal Netherlands Institute for Sea Research (NIOZ), using natural light and seawater. The algae were transported and stored in seawater at room temperature. Measurements on the algae were carried out within 2 days after acquisition. 

\textit{Ulva sp.}, \textit{Undaria pinnatifida},\textit{Grateloupia turuturu}, \textit{Saccharina latissima}, \textit{Fucus serratus} and \textit{Fucus spiralis} were collected by Guido Krijger from WildWier\footnote{Any mention of commercial products or companies within this paper is for information only; it does not imply recommendation or endorsement by the authors or their affiliated institutions.}  from the North Sea near Middelburg in February. The algae were transported under refrigeration and stored in seawater. Measurements on the algae were carried out within 2 days after acquisition. 

Leaves of \textit{Skimmia japonica} and \textit{Prunus laurocerasus} were collected in January from a private backyard garden near the city centre of Amsterdam, \textit{Aspidistra elatior} was obtained from the Hortus Botanicus Vrije Universiteit Amsterdam in February. 

\subsection{Spectropolarimetry}
{\color{red}For all measurements, three different samples were used (n=3) and each single measurement is the average of at least 20000 repetitions.} Before each measurement the samples were padded with paper towels to remove excess surface water. Circular polarization measurements were carried out in transmission and were performed using TreePol. TreePol is a dedicated spectropolarimetric instrument developed by the Astronomical Instrumentation Group at the Leiden Observatory (Leiden University). The instrument was specifically developed to measure the fractional circular polarization ($V/I$) {\color{red}of a sample interacting with unpolarized light} as a function of wavelength (400 nm to 900 nm) and is capable of fast measurements with a sensitivity of $\sim 1*10^{-4}$. {\color{red}TreePol applies spectral multiplexing with the implementation of a dual fiber-fed spectrometer using ferro-liquid-crystal (FLC) modulation synchronized with fast read-out of the one-dimensional detector in each spectrograph, in combination with a dual-beam approach in which a polarizing beam splitter feeds the two spectrographs with orthogonally polarized light (see also \cite{Patty2017}).

In this study we have measured the induced fractional circular polarizance normalized by the total total transmitted light intensity ($V/I$). Circular dichroism measures the differential absorption of left- or right-handed circularly polarized incident light, which is often reported in degrees $\theta$. Under certain conditions, these two can be related and can therefore be converted by $V/I\approx\frac{2\pi\theta_{deg}}{180}$ (see also \cite{Patty2018a}). It has been shown that for leaves in transmission the induced polarizance and the differential absorbance are comparable \cite{Patty2017, Patty2018b}, but we have not verified this for the samples used in this study.} 

\section{Results}
\subsection{Higher Plants}
\begin{figure}[ht]
	\hspace{-1 cm}
	\includegraphics[width=1.1\textwidth]{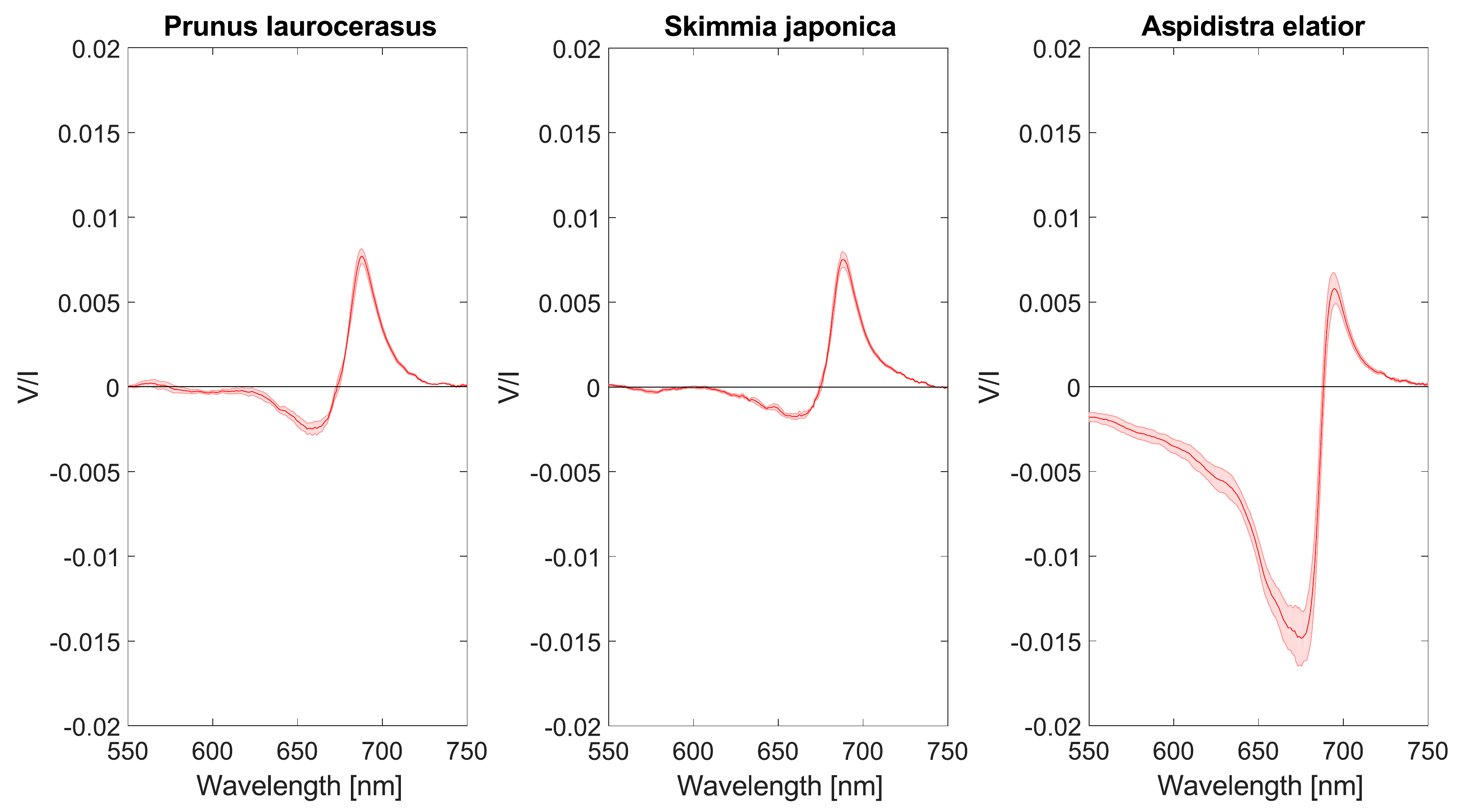}
	\caption{Circular polarimetric spectra of \textit{Skimmia japonica}, \textit{Prunus laurocerasus} and \textit{Aspidistra elatior} leaves. Shaded areas denote the standard error, n=3 per species.}
	\label{fig:Plant}
\end{figure}

The circular polarization spectra of three different higher plants are shown in Figure \ref{fig:Plant}. For all species we observe the typical split signal around the chlorophyll \textit{a} absorption band ($\approx$ 680 nm) with a negative band at $\approx$ 660 nm and a positive band at $\approx$ 690 nm. 
The spectra of \textit{Skimmia} and \textit{Prunus} are very similar to each other in both shape and magnitude and show no significant differences. These results are also very similar to the results obtained for most other higher plants (data not shown). Interestingly, the circular polarimetric spectrum of \textit{Aspidistra elatior} shows an exceedingly large negative band ($-1.5*10^{-2}$) with a noticeable negative circular polarization extending much further into the blue, beyond the chlorophyll \textit{a} (but also \textit{b}) absorption bands. The positive band, however, has a similar magnitude ($+6*10^{-3}$) as the other two plant species.

\subsection{Green algae}
\begin{figure}[ht]
	\includegraphics[width=\textwidth]{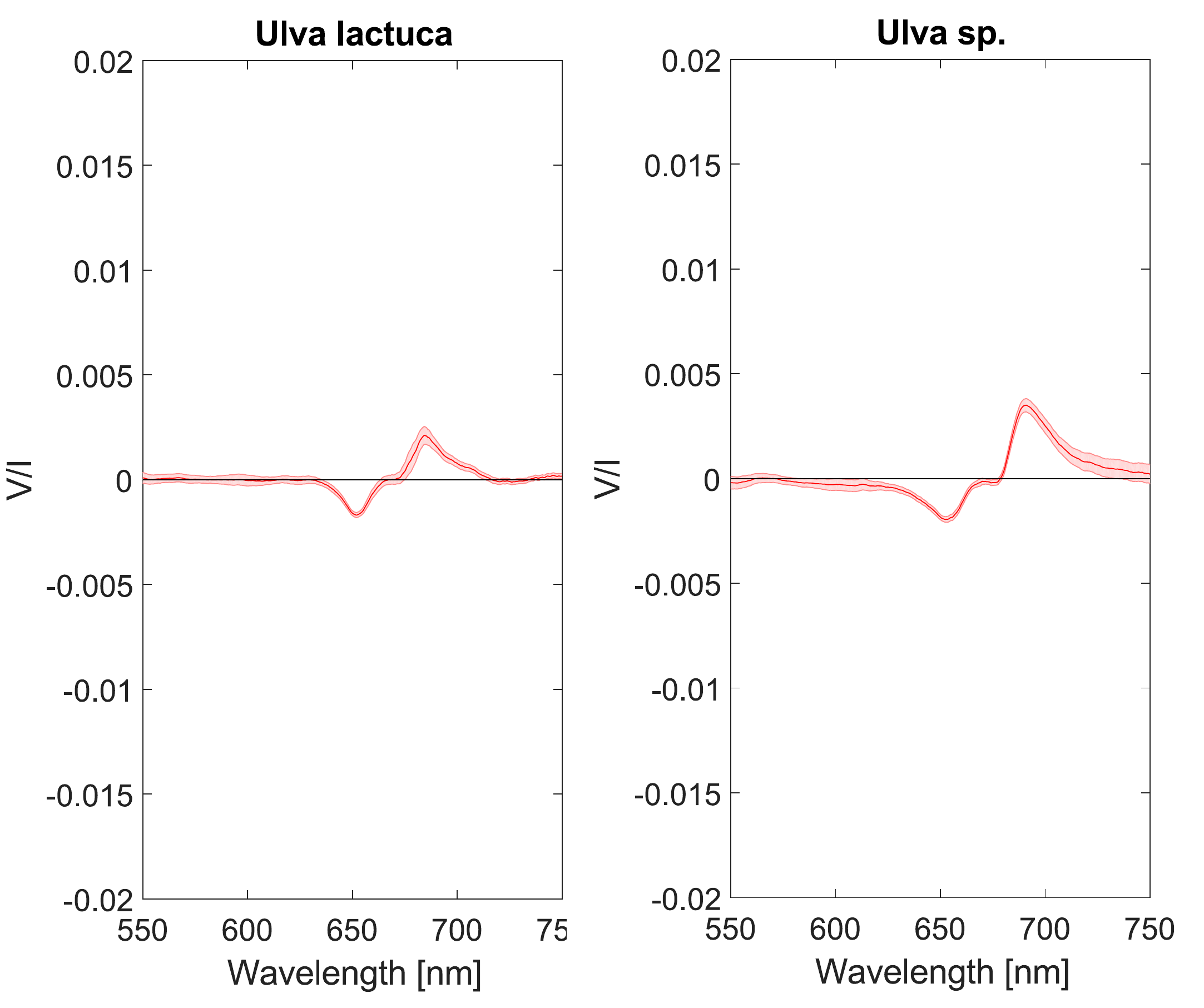}
	\caption{Circular polarimetric spectra of \textit{Ulva lactuca} and \textit{Ulva sp.} green algae. Shaded areas denote the standard error, n=3 per species.}
	\label{fig:Greenalgae}
\end{figure}

The circular polarization spectra of two different green algae are shown in Figure \ref{fig:Greenalgae}. Similar to higher plants a split signal is observed around the chlorophyll \textit{a} absorption band ($\approx$ 680 nm). Unlike higher plants, however, the negative and positive band do not seem to overlap. The negative band reaches a $V/I$ minimum at $\approx$ 655 nm and the positive band reaches a maximum at $\approx$ 690 nm, but the $V/I$ signal is close to 0, and thus shows no net circular polarization between $\approx$ 665 nm to 678 nm. Additionally, the magnitude of the signals is much smaller than that of higher plants.

\subsection{Red algae}
\begin{figure}[h]
	\includegraphics[width=\textwidth]{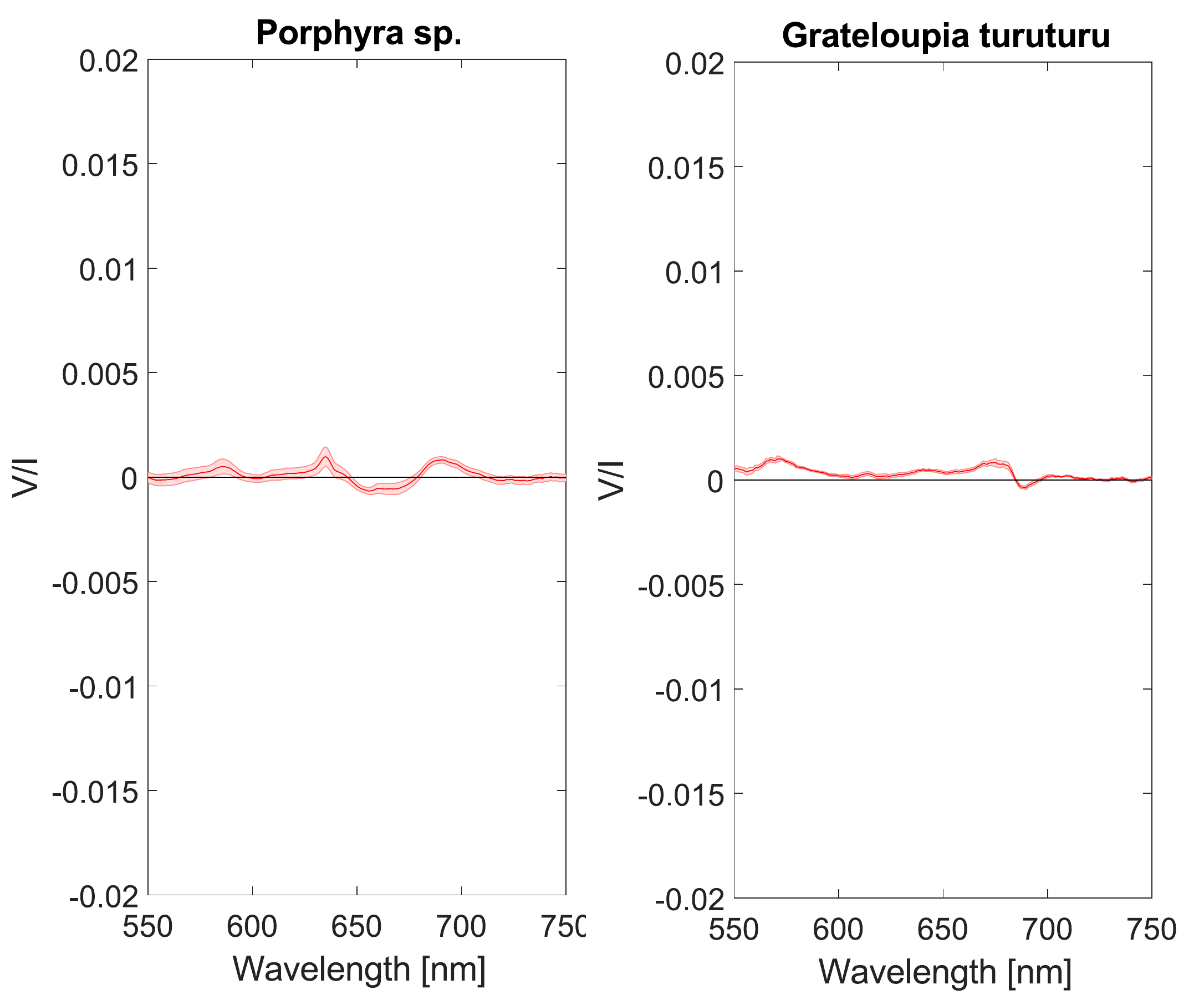}
	\caption{Circular polarimetric spectra of \textit{Porphyra sp.} and \textit{Grateloupia turuturu} red algae. Shaded areas denote the standard error, n=3 per species.}
	\label{fig:Red}
\end{figure}
We show the circular polarization spectra of two different red algae in Figure \ref{fig:Red}. These spectra show distinct differences compared to the higher plants and the green or brown algae. \textit{Porphyra sp.} shows a continuous split signal around $\approx$ 680 nm, and an additional sharp positive feature at $\approx$ 635 nm. \textit{Grateloupia turuturu} lacks these features and shows an inverse split signal around $\approx$ 680 nm. In both species, non-zero circular polarization can also be observed between 550 nm to 600 nm. We will further interpret these results in the Discussion.

\subsection{Brown algae}
\begin{figure}[htb]
	\hspace{-1 cm}
	\includegraphics[width=1.1\textwidth]{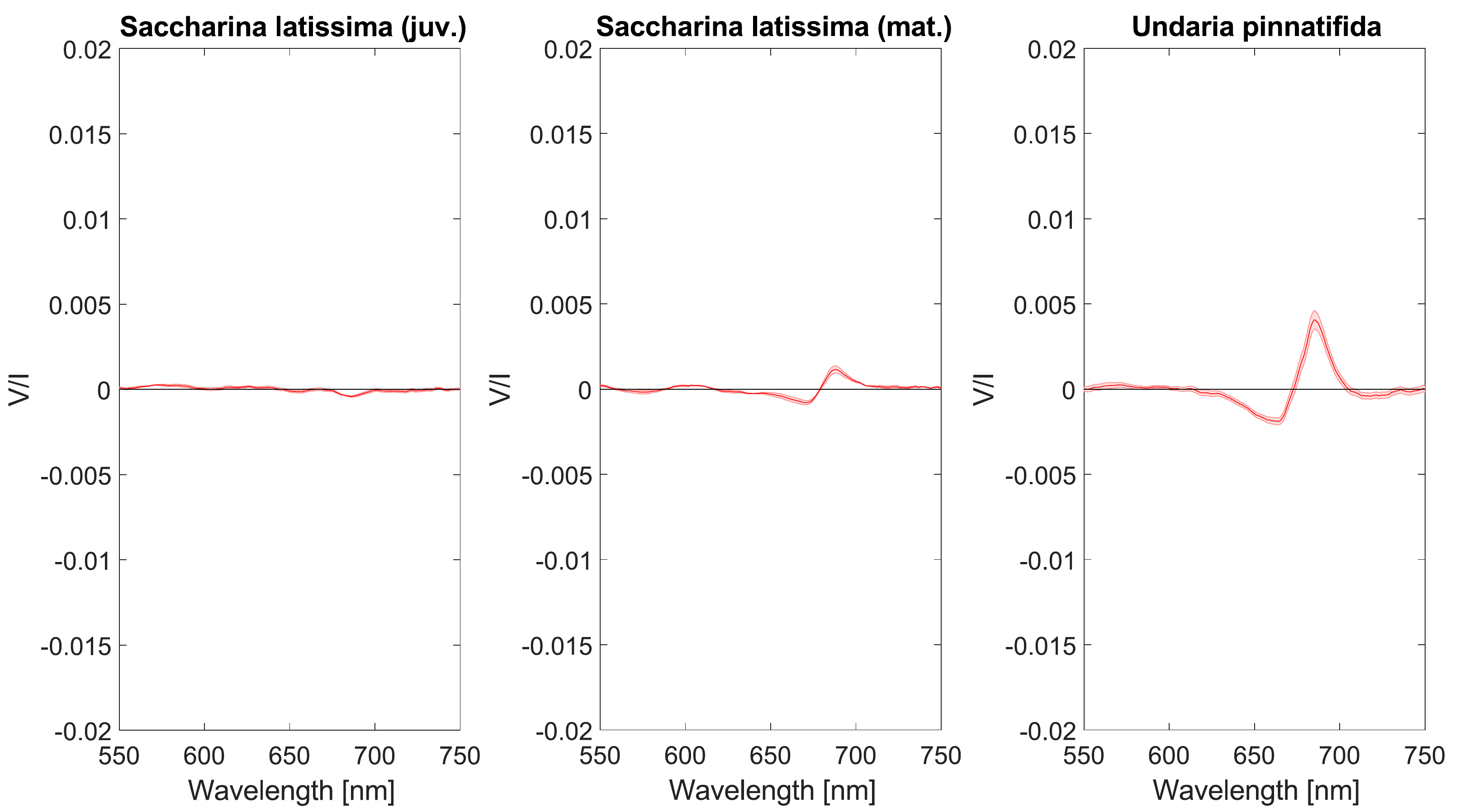}
	\caption{Circular polarimetric spectra of \textit{Saccharina latissima} (juvenile and mature) and \textit{Undaria pinnatifida} brown algae. Shaded areas denote the standard error, n=3 per species.}
	\label{fig:Brown}
\end{figure}
\begin{figure}[h]
	\includegraphics[width=\textwidth]{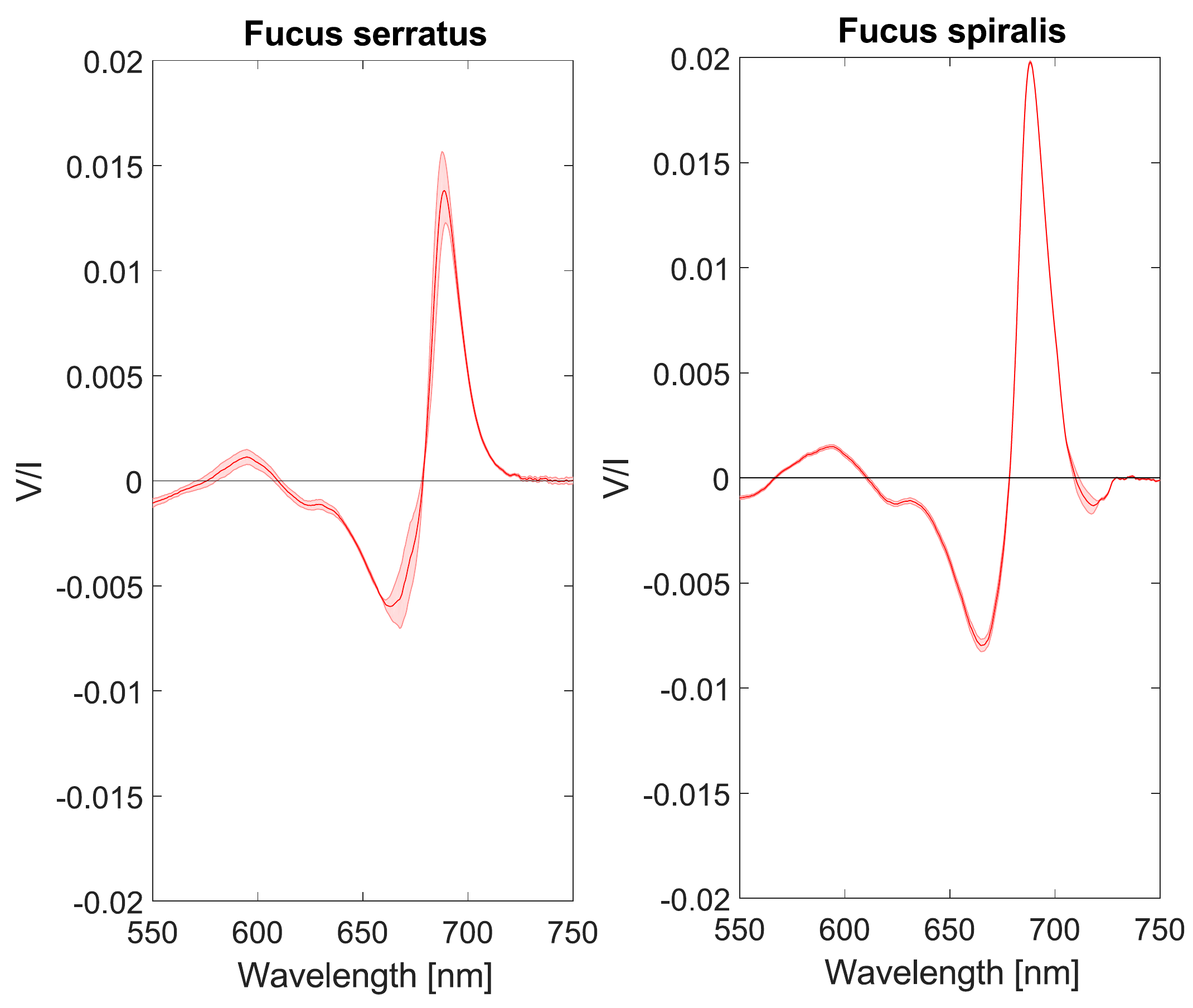}
	\caption{Circular polarimetric spectra of \textit{Fucus serratus} and \textit{Fucus spiralis} brown algae. Shaded areas denote the standard error, n=3 per species.}
	\label{fig:Fucus}
\end{figure}

The brown algae exhibit a lot of variation in signal strength. For ease of comparison, the results of our circular spectropolarimetric measurements are plotted in Figures \ref{fig:Brown} and \ref{fig:Fucus} on the same y-scale. Figure \ref{fig:Brown} makes clear that a juvenile \textit{Saccharina latissima} barely displays a significant signal with the exception of a very weak negative feature ($V/I = -4*10^{-4}$). The mature \textit{Saccharina latissima} samples show somewhat stronger bands, although the signal is still relatively small ($-1*10^{-3}, +1*10^{-3}$). The polarimetric spectra of the brown algae \textit{Undaria pinnatifida}, displays a larger signal comparable to that of higher vegetation. 

Interestingly, the polarimetric spectra of the brown algae of the genus \textit{Fucus} displays very large circular polarization signals, see Figure \ref{fig:Fucus}. The alga \textit{Fucus spiralis} has a $V/I$ minimum and maximum of $-8*10^{-3}$ and $+2*10^{-2}$ respectively. Additionally, the bands are relatively narrow, with less polarization outside of the chlorophyll \textit{a} absorbance band. In the polarimetric spectra of \textit{Fucus spiralis}, and to a lesser extent also of \textit{Undaria pinnatifida}, a small negative band can be observed at 720 nm. Additionally, in the spectra of both \textit{Fucus serratus} and \textit{Fucus spiralis} a positive band can be observed at 595 nm.

\begin{figure}[h]
	\includegraphics[width=\textwidth]{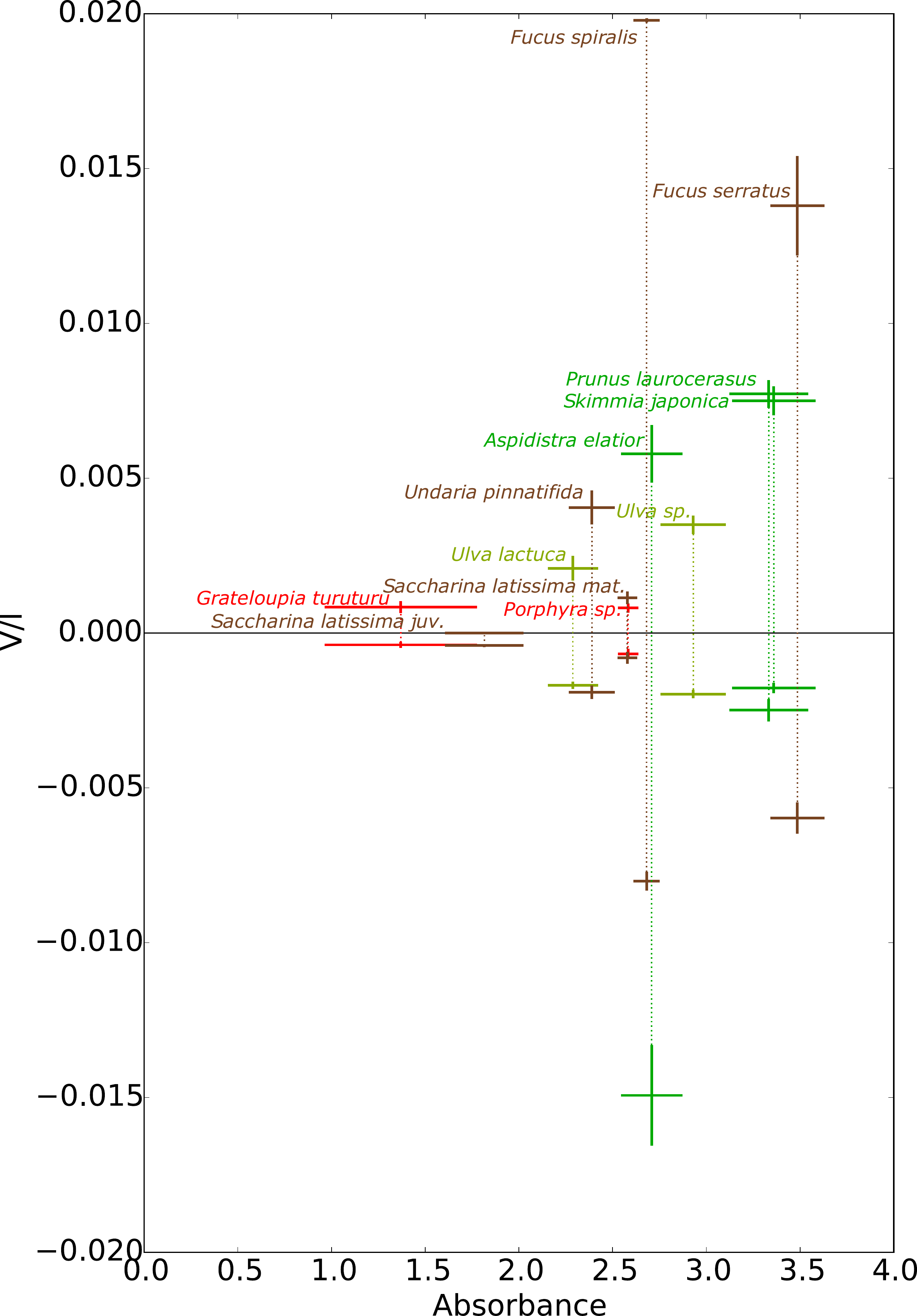}
	\caption{Maximum extend of the V/I bands within 650 nm - 700 nm  against the absorbance over 675 nm - 685 nm. Error bars denote the standard error for n=3 per species}
	\label{fig:VIabs}
\end{figure}

 {\color{red}\subsection{$V/I$ versus absorbance}

The $V/I$ maxima and minima versus the absorbance are shown in Figure \ref{fig:VIabs}. A slight correlation is visible between the maximum and minimum magnitude of the $V/I$ bands within 650 nm to 700 nm and the absorbance over 675 nm to 685 nm. In general, the magnitude of the bands increases with increasing absorbance. Both \textit{Fucus serratus} and \textit{Fucus spiralis} show positive and negative bands with a very large magnitude well outside this trend. This is similar for the large negative band of \textit{Aspidistra elatior}. On the other hand, mature \textit{Saccharina latissima} and \textit{Porphyra sp.} have a relatively low circular polarizance.}

\clearpage

\section{Discussion}
\begin{figure}[h]
	\includegraphics[width=\textwidth]{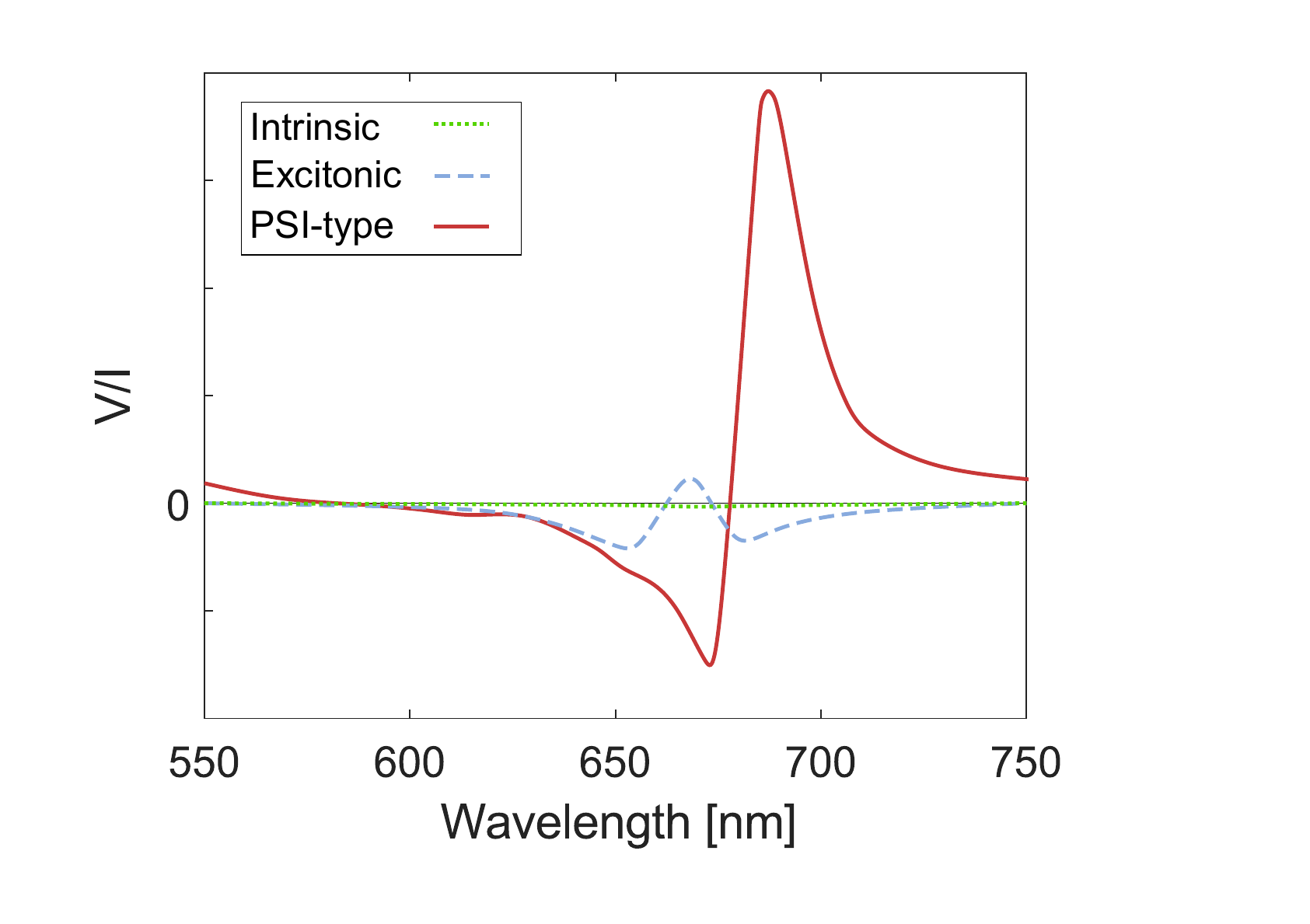}
	\caption{The three major sources of circular polarizance around the chlorophyll absorbance band in the red for higher plants for identical chlorophyll concentrations. Adapted after \cite{Garab2009}. }
	\label{fig:Psi}
\end{figure}

Different eukaryotic phototrophic organisms display different circular polarization spectra, with signal magnitudes that can vary by two orders of magnitude. Chlorophyll \textit{a} itself exhibits a very weak intrinsic circular polarizance around 680 nm\cite{Garab2009}. Excitonic coupling between chlorophylls leads to a much larger signal in phototrophic bacteria and certain algae. In many more developed phototrophic organisms the polarization spectra are dominated by the density and handedness of the supramolecular structures (psi-type circular dichroism), although these signals are superimposed on each other. {\color{red} Thus for identical chlorophyll concentrations the polarimetric spectral characteristics can vastly differ depending on the organization (see also Figure \ref{fig:Psi}).}

The typical psi-type circular spectropolarimetric signals observed in vegetation are the result of the superposition of two relatively independent signals resulting from different chiral macrodomains in the chloroplast \cite{Garab1988a, Garab1988b, Finzi1989, Garab1991}. These psi-type bands of opposite sign do not have the same spectral shape and thus do not cancel each other out completely. The negative band is predominantly associated with the stacking of the thylakoid membranes, whereas the positive band mainly derives from the lateral organization of the chiral macrodomains formed by the PSII-LHCII complexes \cite{Cseh2000, Dobrikova2003, Jajoo2012, Garab1991}. 

Plant chloroplasts generally show little variation in structure \cite{Staehelin1986}, which is noticeable in the circular polarization spectra of most plants (e.g. see the spectra of \textit{Skimmia} and \textit{Prunus} in Figure \ref{fig:Plant}). It has been reported before that the cpDNA sequences are extraordinarily conserved among plants and nearly identical in ferns, gymnosperms and angiosperms \cite{Palmer1986}. Of course, certain plants contain more chloroplasts per cell, or contain chloroplasts which are significantly larger or smaller, but in both cases the normalized circular polarization will remain the same. 

The polarimetric spectra of \textit{Aspidistra} (Figure \ref{fig:Plant}) show a remarkably intense negative band, unlike the results usually encountered in plants. {\color{red}The positive band, however, has a magnitude that can be expected based on the lower absorbance as compared to the other higher plants we measured (see also Figure \ref{fig:VIabs}).} It has been shown that the contribution of both the negative and the positive band is dependent on the alignment of the chloroplasts \cite{Garab1991, Garab1988b}, which might locally be aligned in such a way that only a single band dominates (e.g. near the veins of leaves \cite{Patty2018b}). The polarimetric spectra of \textit{Aspidistra}, however, can be very well explained by the unusually large grana. Previous electron microscopy research on \textit{Aspidistra elatior} chloroplasts revealed grana containing a vast number of thylakoid layers that may well exceed 100 \cite{Steinmann1955}. As the positive and the negative bands overlap (leading to the split signal), it is to be expected that also the positive band is larger than encountered normally.

Similar to higher plants, also green algae contain PSII-LHCII supercomplexes utilized in photosynthesis. Between green algae and higher plants there are slight differences in the trimeric LHCII proteins and their isoforms, and, in addition, the green algae lack one of three minor monomeric LHCII polypeptides (CP24) (see also\cite{Minagawa2013} and references therein). The green algae we measured show a spectral polarimetric profile that appears very similar to that of plants. However, the negative band centered around 650 nm is likely an excitonic band resulting from short-range interactions of the chlorophylls and the negative, usually stronger, psi-type band around 675 nm is virtually absent. The positive psi-type centered around 690 nm, on the other hand, is still present.

These results are unlike those reported for the unicellular green algae \textit{Chlamydomonas reinhardtii}, which display a negative excitonic and a negative psi-type band of equal strength (e.g. see \cite{Nagy2014}). Importantly, the PSII-LHCII supercomplexes are far less stable in green algae as compared to plants, and it has been indicated that the L trimer (as well as the M and S trimers) could dissociate easily from PSII \cite{Tokutsu2012}. It has been shown that in \textit{Ulva} flattening of the chloroplasts occurs under illumination, which additionally results in a decrease in thickness of the thylakoid membrane itself \cite{Murakami1970}. Such fundamental changes in molecular structure might easily lead to (partial) dissociation of trimer L, which in turn can lead to the observed apparent absence of the negative psi-type band. 

The red algae contain a more primitive photosynthetic apparatus that represents a transition between cyanobacteria and the chloroplasts of other algae and plants. This is also very evident from the displayed spectra in Figure \ref{fig:Red}. For both species the magnitude of the signal is small {\color{red}and comparable, even though \textit{Porphyra sp.} had a much larger absorbance (Figure \ref{fig:VIabs}),} but the spectral shape suggest very fundamental differences in molecular structure. Surprisingly, \textit{Porphyra sp.} shows a circular polarization spectrum with bands that might be associated with psi-type circular polarizance (at 675 nm (-) and at 690 nm (+)). The origin and significance of these signals, however, requires further investigation. The circular polarimetric spectra of \textit{Grateloupia turuturu} lack these features but show two bands that can be associated with the excitonic circular polarization bands similar to those in cyanobacteria (at 670 nm (+) and at 685 nm (-)) (cf. \cite{Sparks2009}), which for a large part result from the excitonic interactions in PSI \cite{Schlodder2007}. In both species, the features between 550 nm and 600 nm might be associated with R-phycoerythrin \cite{Bekasova2013}. Additionally, in \textit{Porphyra sp.} the sharp feature around 635 nm can be associated with phycocyanin \cite{Sparks2009}. Both pigment-protein complexes belong to the phycobilisomes, which only occur in red algae and cyanobacteria and function as light harvesting antennae for PSII while LHC is limited to PSI. 

As in red algae and green algae, the brown algae contain no true grana but the thylakoid membranes are stacked in groups of three \cite{Berkaloff1983}. The brown algae measured in this study additionally contain chlorophyll \textit{c}, which is slightly blue-shifted compared to chlorophyll \textit{a} or \textit{b}. Compared to chlorophyll \textit{a}, chlorophyll \textit{c} however has only a very weak contribution to the overall circular polarizance. Additionally, in brown algae the light-harvesting antennae are homogeneously distributed along the thylakoid membranes \cite{DeMartino2000, Buechel1997}. 

Interestingly, the juvenile \textit{Saccharina} displays only a very weak negative band around 683 nm (Figure \ref{fig:Brown}). These results closely resemble those of isolated brown algae LHCs, which exhibit no excitonic bands but show solely a negative band around 680 nm. This band likely results from an intrinsic induced chirality of the chlorophyll \textit{a} protein complex \cite{Buechel1997}. The polarimetric spectra of mature \textit{Saccharina} and \textit{Undaria} show a split signal that is similar to that of higher plants. While the molecular architecture of the LHCs is very different from those in higher plants, the pigment-protein complexes in brown algae are organized in large chiral domains which give similar psi-type signals in circular polarizance \cite{Szabo2008, Nagy2012}. These intrinsic so-called fucoxanthin chlorophyll \textit{a/c} binding proteins show a high homology to LHC in higher plants and have been shown to form complexes with trimers or higher oligomers \cite{Lepetit2007, Buechel2003, Katoh1989}.

As shown in Figure \ref{fig:Fucus}, the measured species of the genus \textit{Fucus} exhibit an unusually large signal in circular polarizance, {\color{red}while the absorbance of the samples was within the range of the samples of the other species (Figure \ref{fig:VIabs})}. Although their spectral shapes are very similar to those of diatoms (cf. \cite{Ghazaryan2016, Szabo2008, Buechel1997}) the bands are 2 orders of magnitude stronger in \textit{Fucus}. Most research on chlorophyll \textit{a/c} photosynthesis is, however, carried out on diatoms and the reported size of the protein complexes again varies. Signals of such magnitude suggest that  these macromolecular assemblies are much larger in \textit{Fucus} than previously reported for other brown algae. Additionally, in the spectra of \textit{Fucus} a positive band can be observed around 595 nm. Most likely, this band and the weaker negative band around 625 nm can be assigned to chlorophyll \textit{c}. 

The results here show that the molecular and macromolecular organization of the photosynthetic machinery in algae is much more flexible and dynamic than reported before, likely due to larger inter-specific differences than generally assumed. Additionally, this also appears to be the case for one of the plants we measured (\textit{Aspidistra elatior}), which displayed a negative psi-band one order of magnitude larger than ordinarily observed for higher plants. 

When it comes to circular polarizance as a biosignature, it is important to note that efficient photosynthesis is not necessarily accompanied by large signals in circular polarization. While the intrinsic circular polarizance of chlorophyll is very low, the magnitude of the signals become greatly enhanced by a larger organization resulting in excitonic circular polarizance and ultimately psi-type circular polarizance. For the latter, the chiral organization of the macrodomains of the pigment-protein complexes is of importance, but it should be noted that the density of the complexes needs to be large enough (that is, significant coupling over the macrodomain is required) in order to function as a chiral macrodomain \cite{Keller1986}. Many organisms thus display only excitonic circular polarizance, as is the case for certain algae measured in this study and generally bacteria. When psi-type circular polarizance is possible, the signals can easily become very large, in our study up to 2 $\%$ for brown algae in transmission. 

\subsection{Conclusions}

We have measured the polarizance of various multicellular algae representing different evolutionary stages of eukaryotic photosynthesis. We have shown that the chiral organization of the macrodomains can vary greatly between these species. Future studies using molecular techniques to further characterize and isolate the complexes in these organisms are highly recommended. It will additionally prove very interesting to investigate these chloroplasts (including those with larger grana such as \textit{Aspidistra)} using polarization microscopy (e.g. \cite{Steinbach2014, Finzi1989, Gombos2008}). The high quality spectra in this study and their reproducibility underline the possibility of utilizing polarization spectroscopy as a quantitative tool for non-destructively probing the molecular architecture \textit{in vivo}.

Our results not only show variations in spectral shapes, but also in magnitude. Especially the brown algae show a large variation, which is up to three orders of magnitude for the species measured in this study. Additionally, the induced fractional circular polarization by members of the genus \textit{Fucus} is much larger than observed in vegetation. Future studies on the supramolecular organization in this genus and the variability caused by, for instance, light conditions, will also clarify the maximum extent of the circular polarizance by oxygenic photosynthetic organisms. 

While the displayed results were obtained in transmission, the spectral features are also present in reflection. As such, future use of circular spectropolarimetry in satellite or airborne remote sensing could aid in detecting the presence of floating multicellular algae but also aid in species differentiation, which is important in regional biogeochemistry \cite{Dierssen2015}.

Importantly, while the presence of similar circular polarization signals is an unambiguous indicator for the presence of life, life might also flourish on a planetary surface and still show minimal circular polarizance (which for instance would have been the case on Earth if terrestrial vegetation evolved through different Archaeplastida/SAR supergroup lineages). On the other hand, these signals might also be much larger than we would observe from an Earth disk-averaged spectrum {\color{red}(which is the unresolved and therefore spatially integrated spectrum of a planet)}. 

\section{Acknowledgments}
We thank Klaas Timmermans (NIOZ) and Guido Krijger (Wildwier) for providing us with the algae samples. We thank the Hortus Botanicus Vrije Universiteit Amsterdam for providing us with the Aspidistra samples. This work was supported by the Planetary and Exoplanetary Science Programme (PEPSci), grant 648.001.004, of the Netherlands Organisation for Scientific Research (NWO). 
\nolinenumbers
\clearpage

\bibliography{./Library/Alles}

\begin{thebibliography}{10}

\bibitem{Schwieterman2017}
E.~W. Schwieterman, N.~Y. Kiang, M.~N. Parenteau, C.~E. Harman, S.~DasSarma,
  T.~M. Fisher, G.~N. Arney, H.~E. Hartnett, C.~T. Reinhard, S.~L. Olson, V.~S.
  Meadows, C.~S. Cockell, S.~I. Walker, J.~L. Grenfell, S.~Hegde, S.~Rugheimer,
  R.~Hu, and T.~W. Lyons, ``Exoplanet biosignatures: A review of remotely
  detectable signs of life,'' {\em Astrobiology}, 2018.

\bibitem{Fasman2013}
G.~D. Fasman, {\em Circular dichroism and the conformational analysis of
  biomolecules}.
\newblock Springer Science \& Business Media, 2013.

\bibitem{Patty2018a}
C.~H.~L. Patty, I.~L. ten Kate, W.~B. Sparks, and F.~Snik, ``Remote sensing of
  homochirality: a proxy for the detection of extraterrestrial life,'' in {\em
  Chiral Analysis, Second Edition: Advances in Spectroscopy, Chromatography and
  Emerging Methods} (P.~Polavarapu, ed.), ch.~2, pp.~29--69, Elsevier Science,
  2018.

\bibitem{Garab2009}
G.~Garab and H.~van Amerongen, ``Linear dichroism and circular dichroism in
  photosynthesis research,'' {\em Photosynthesis Research}, vol.~101, no.~2-3,
  pp.~135--146, 2009.

\bibitem{Keller1986}
D.~Keller and C.~Bustamante, ``Theory of the interaction of light with large
  inhomogeneous molecular aggregates. {II.} psi-type circular dichroism,'' {\em
  The Journal of Chemical Physics}, vol.~84, no.~6, pp.~2972--2980, 1986.

\bibitem{Garab1991}
G.~Garab, L.~Finzi, and C.~Bustamante, ``Differential polarization imaging of
  chloroplasts: Microscopic and macroscopic linear and circular dichroism,''
  {\em Light in Biology and Medicine}, vol.~2, 1991.

\bibitem{TinocoJr1987}
I.~Tinoco~Jr, W.~Mickols, M.~Maestre, and C.~Bustamante, ``Absorption,
  scattering, and imaging of biomolecular structures with polarized light,''
  {\em Annual Review of Biophysics and Biophysical Chemistry}, vol.~16, no.~1,
  pp.~319--349, 1987.

\bibitem{Patty2017}
C.~H.~L. Patty, L.~J.~J. Visser, F.~Ariese, W.~J. Buma, W.~B. Sparks, R.~J.~M.
  van Spanning, W.~F.~M. Röling, and F.~Snik, ``Circular spectropolarimetric
  sensing of chiral photosystems in decaying leaves,'' {\em Journal of
  Quantitative Spectroscopy and Radiative Transfer}, vol.~189, pp.~303--311,
  2017.

\bibitem{Patty2018b}
C.~H.~L. Patty, D.~A. Luo, F.~Snik, F.~Ariese, W.~J. Buma, I.~L. ten Kate,
  R.~J.~M. van Spanning, W.~B. Sparks, T.~A. Germer, G.~Garab, and M.~W.
  Kudenov, ``Imaging linear and circular polarization features in leaves with
  complete {Mueller} matrix polarimetry,'' {\em Biochimica et Biophysica Acta
  ({BBA}) - General Subjects}, vol.~1862, no.~6, 2018.

\bibitem{Sparks2009}
W.~B. Sparks, J.~Hough, T.~A. Germer, F.~Chen, S.~DasSarma, P.~DasSarma, F.~T.
  Robb, N.~Manset, L.~Kolokolova, N.~Reid, {\em et~al.}, ``Detection of
  circular polarization in light scattered from photosynthetic microbes,'' {\em
  Proceedings of the National Academy of Sciences}, vol.~106, no.~19,
  pp.~7816--7821, 2009.

\bibitem{Sparks2009a}
W.~Sparks, J.~Hough, L.~Kolokolova, T.~Germer, F.~Chen, S.~DasSarma,
  P.~DasSarma, F.~Robb, N.~Manset, I.~Reid, F.~Macchetto, and W.~Martin,
  ``Circular polarization in scattered light as a possible biomarker,'' {\em
  Journal of Quantitative Spectroscopy and Radiative Transfer}, vol.~110,
  no.~14-16, pp.~1771--1779, 2009.

\bibitem{Wolstencroft1974}
R.~D. Wolstencroft, ``The circular polarization of light reflected from certain
  optically active surfaces,'' in {\em IAU Colloq. 23: Planets, Stars, and
  Nebulae: Studied with Photopolarimetry}, p.~495, 1974.

\bibitem{Pospergelis1969}
M.~Pospergelis, ``Spectroscopic measurements of the four stokes parameters for
  light scattered by natural objects.,'' {\em Soviet Astronomy}, vol.~12,
  p.~973, 1969.

\bibitem{Heckman2001}
D.~S. Heckman, D.~M. Geiser, B.~R. Eidell, R.~L. Stauffer, N.~L. Kardos, and
  S.~B. Hedges, ``Molecular evidence for the early colonization of land by
  fungi and plants,'' {\em Science}, vol.~293, no.~5532, pp.~1129--1133, 2001.

\bibitem{Wellman2000}
C.~H. Wellman and J.~Gray, ``The microfossil record of early land plants,''
  {\em Philosophical Transactions of the Royal Society of London B: Biological
  Sciences}, vol.~355, no.~1398, pp.~717--732, 2000.

\bibitem{Nutman2016}
A.~P. Nutman, V.~C. Bennett, C.~R. Friend, M.~J. Van~Kranendonk, and A.~R.
  Chivas, ``Rapid emergence of life shown by discovery of
  3,700-million-year-old microbial structures,'' {\em Nature}, vol.~537,
  no.~7621, pp.~535--538, 2016.

\bibitem{Planavsky2014}
N.~Planavsky, D.~Asael, A.~Hofmann, C.~Reinhard, S.~Lalonde, A.~Knudsen,
  X.~Wang, F.~Ossa, E.~Pecoits, A.~Smith, {\em et~al.}, ``Evidence for oxygenic
  photosynthesis half a billion years before the great oxidation event: Nature
  geoscience, v. 7,'' {\em doi}, vol.~10, pp.~283--286, 2014.

\bibitem{Moreira2000}
D.~Moreira, H.~Le~Guyader, and H.~Philippe, ``The origin of red algae and the
  evolution of chloroplasts,'' {\em Nature}, vol.~405, no.~6782, pp.~69--72,
  2000.

\bibitem{Ponce-Toledo2017}
R.~I. Ponce-Toledo, P.~Deschamps, P.~L{\'o}pez-Garc{\'\i}a, Y.~Zivanovic,
  K.~Benzerara, and D.~Moreira, ``An early-branching freshwater cyanobacterium
  at the origin of plastids,'' {\em Current Biology}, vol.~27, no.~3,
  pp.~386--391, 2017.

\bibitem{McFadden2001}
G.~I. McFadden, ``Primary and secondary endosymbiosis and the origin of
  plastids,'' {\em Journal of Phycology}, vol.~37, no.~6, pp.~951--959, 2001.

\bibitem{Green2011}
B.~R. Green, ``Chloroplast genomes of photosynthetic eukaryotes,'' {\em The
  plant journal}, vol.~66, no.~1, pp.~34--44, 2011.

\bibitem{Andersen1992}
R.~Andersen, ``Diversity of eukaryotic algae,'' {\em Biodiversity \&
  Conservation}, vol.~1, no.~4, pp.~267--292, 1992.

\bibitem{Bengtson2017}
S.~Bengtson, T.~Sallstedt, V.~Belivanova, and M.~Whitehouse,
  ``Three-dimensional preservation of cellular and subcellular structures
  suggests 1.6 billion-year-old crown-group red algae,'' {\em PLoS biology},
  vol.~15, no.~3, p.~e2000735, 2017.

\bibitem{Mustardy2003}
L.~Must{\'a}rdy and G.~Garab, ``Granum revisited. a three-dimensional
  model--where things fall into place,'' {\em Trends in Plant Science}, vol.~8,
  no.~3, pp.~117--122, 2003.

\bibitem{Anderson1973}
J.~M. Anderson, D.~Goodchild, and N.~Boardman, ``Composition of the
  photosystems and chloroplast structure in extreme shade plants,'' {\em
  Biochimica et Biophysica Acta (BBA)-Bioenergetics}, vol.~325, no.~3,
  pp.~573--585, 1973.

\bibitem{Steinmann1955}
E.~Steinmann and F.~Sj{\"o}strand, ``The ultrastructure of chloroplasts,'' {\em
  Experimental cell research}, vol.~8, no.~1, pp.~15--23, 1955.

\bibitem{Faludi-Daniel1973}
A.~Faludi-Daniel, S.~Demeter, and A.~S. Garay, ``Circular dichroism spectra of
  granal and agranal chloroplasts of maize,'' {\em Plant Physiology}, vol.~52,
  no.~1, pp.~54--56, 1973.

\bibitem{Boekema1999}
E.~J. Boekema, H.~van Roon, J.~F.~L. van Breemen, and J.~P. Dekker,
  ``Supramolecular organization of photosystem {II} and its light-harvesting
  antenna in partially solubilized photosystem {II} membranes,'' {\em European
  Journal of Biochemistry}, vol.~266, no.~2, pp.~444--452, 1999.

\bibitem{Toth2016}
T.~N. T{\'o}th, N.~Rai, K.~Solymosi, O.~Zsiros, W.~P. Schr{\"o}der, G.~Garab,
  H.~van Amerongen, P.~Horton, and L.~Kov{\'a}cs, ``Fingerprinting the
  macro-organisation of pigment-protein complexes in plant thylakoid membranes
  in vivo by circular-dichroism spectroscopy,'' {\em Biochimica et Biophysica
  Acta (BBA)-Bioenergetics}, vol.~1857, no.~9, pp.~1479--1489, 2016.

\bibitem{Garab1988}
G.~Garab, R.~C. Leegood, D.~A. Walker, J.~C. Sutherland, and G.~Hind,
  ``Reversible changes in macroorganization of the light-harvesting chlorophyll
  a/b pigment-protein complex detected by circular dichroism,'' {\em
  Biochemistry}, vol.~27, no.~7, pp.~2430--2434, 1988.

\bibitem{Garab1991a}
G.~Garab, J.~Kieleczawa, J.~C. Sutherland, C.~Bustamante, and G.~Hind,
  ``Organization of pigment-protein complexes into macrodomains in the
  thylakoid membranes of wild-type and chlorophyll fo-less mutant of barley as
  revealed by circular dichroism,'' {\em Photochemistry and Photobiology},
  vol.~54, no.~2, pp.~273--281, 1991.

\bibitem{Cseh2000}
Z.~Cseh, S.~Rajagopal, T.~Tsonev, M.~Busheva, E.~Papp, and G.~Garab,
  ``Thermooptic effect in chloroplast thylakoid membranes. thermal and light
  stability of pigment arrays with different levels of structural complexity,''
  {\em Biochemistry}, vol.~39, no.~49, pp.~15250--15257, 2000.

\bibitem{Albertsson2001}
P.-{\AA}. Albertsson, ``A quantitative model of the domain structure of the
  photosynthetic membrane,'' {\em Trends in Plant Science}, vol.~6, no.~8,
  pp.~349--354, 2001.

\bibitem{Nevo2012}
R.~Nevo, D.~Charuvi, O.~Tsabari, and Z.~Reich, ``Composition, architecture and
  dynamics of the photosynthetic apparatus in higher plants,'' {\em The Plant
  Journal}, vol.~70, no.~1, pp.~157--176, 2012.

\bibitem{Trissl1993}
H.-W. Trissl and C.~Wilhelm, ``Why do thylakoid membranes from higher plants
  form grana stacks?,'' {\em Trends in Biochemical Sciences}, vol.~18, no.~11,
  pp.~415--419, 1993.

\bibitem{Anderson1999}
J.~M. Anderson, ``Insights into the consequences of grana stacking of thylakoid
  membranes in vascular plants: a personal perspective,'' {\em Functional Plant
  Biology}, vol.~26, no.~7, pp.~625--639, 1999.

\bibitem{Collen2013}
J.~Collen, B.~Porcel, W.~Carre, S.~G. Ball, C.~Chaparro, T.~Tonon,
  T.~Barbeyron, G.~Michel, B.~Noel, K.~Valentin, M.~Elias, F.~Artiguenave,
  A.~Arun, J.-M. Aury, J.~F. Barbosa-Neto, J.~H. Bothwell, F.-Y. Bouget,
  L.~Brillet, F.~Cabello-Hurtado, S.~Capella-Gutierrez, B.~Charrier,
  L.~Cladiere, J.~M. Cock, S.~M. Coelho, C.~Colleoni, M.~Czjzek, C.~D. Silva,
  L.~Delage, F.~Denoeud, P.~Deschamps, S.~M. Dittami, T.~Gabaldon, C.~M.~M.
  Gachon, A.~Groisillier, C.~Herve, K.~Jabbari, M.~Katinka, B.~Kloareg,
  N.~Kowalczyk, K.~Labadie, C.~Leblanc, P.~J. Lopez, D.~H. McLachlan,
  L.~Meslet-Cladiere, A.~Moustafa, Z.~Nehr, P.~N. Collen, O.~Panaud,
  F.~Partensky, J.~Poulain, S.~A. Rensing, S.~Rousvoal, G.~Samson,
  A.~Symeonidi, J.~Weissenbach, A.~Zambounis, P.~Wincker, and C.~Boyen,
  ``Genome structure and metabolic features in the red seaweed {Chondrus}
  crispus shed light on evolution of the archaeplastida,'' {\em Proceedings of
  the National Academy of Sciences}, vol.~110, no.~13, pp.~5247--5252, 2013.

\bibitem{Tokutsu2012}
R.~Tokutsu, N.~Kato, K.~H. Bui, T.~Ishikawa, and J.~Minagawa, ``Revisiting the
  supramolecular organization of photosystem {II} in {Chlamydomonas}
  reinhardtii,'' {\em Journal of Biological Chemistry}, vol.~287, no.~37,
  pp.~31574--31581, 2012.

\bibitem{Remias2005}
D.~Remias, U.~L{\"u}tz-Meindl, and C.~L{\"u}tz, ``Photosynthesis, pigments and
  ultrastructure of the alpine snow alga {Chlamydomonas} nivalis,'' {\em
  European Journal of Phycology}, vol.~40, no.~3, pp.~259--268, 2005.

\bibitem{Gunning1999}
B.~Gunning and O.~Schwartz, ``Confocal microscopy of thylakoid autofluorescence
  in relation to origin of grana and phylogeny in the green algae,'' {\em
  Functional Plant Biology}, vol.~26, no.~7, pp.~695--708, 1999.

\bibitem{Larkum2003}
A.~W. Larkum and M.~Vesk, ``Algal plastids: Their fine structure and
  properties,'' in {\em Photosynthesis in Algae}, pp.~11--28, Springer
  Netherlands, 2003.

\bibitem{Larkum2005}
A.~W. Larkum and M.~Kühl, ``Chlorophyll d: the puzzle resolved,'' {\em Trends
  in Plant Science}, vol.~10, no.~8, pp.~355--357, 2005.

\bibitem{Premvardhan2010}
L.~Premvardhan, B.~Robert, A.~Beer, and C.~B{\"u}chel, ``Pigment organization
  in fucoxanthin chlorophyll a/c2 proteins (fcp) based on resonance raman
  spectroscopy and sequence analysis,'' {\em Biochimica et Biophysica Acta
  (BBA)-Bioenergetics}, vol.~1797, no.~9, pp.~1647--1656, 2010.

\bibitem{Buechel2015}
C.~B{\"u}chel, ``Evolution and function of light harvesting proteins,'' {\em
  Journal of Plant Physiology}, vol.~172, pp.~62--75, 2015.

\bibitem{Burki2016}
F.~Burki, M.~Kaplan, D.~V. Tikhonenkov, V.~Zlatogursky, B.~Q. Minh, L.~V.
  Radaykina, A.~Smirnov, A.~P. Mylnikov, and P.~J. Keeling, ``Untangling the
  early diversification of eukaryotes: a phylogenomic study of the evolutionary
  origins of {Centrohelida}, {Haptophyta} and {Cryptista},'' in {\em Proc. R.
  Soc. B}, vol.~283, p.~20152802, The Royal Society, The Royal Society, jan
  2016.

\bibitem{Dorrell2011}
R.~G. Dorrell and A.~G. Smith, ``Do red and green make brown?: perspectives on
  plastid acquisitions within chromalveolates,'' {\em Eukaryotic Cell},
  vol.~10, no.~7, pp.~856--868, 2011.

\bibitem{Garab1988a}
G.~Garab, S.~Wells, L.~Finzi, and C.~Bustamante, ``Helically organized
  macroaggregates of pigment-protein complexes in chloroplasts: evidence from
  circular intensity differential scattering,'' {\em Biochemistry}, vol.~27,
  no.~16, pp.~5839--5843, 1988.

\bibitem{Garab1988b}
G.~Garab, A.~Faludi-Daniel, J.~C. Sutherland, and G.~Hind, ``Macroorganization
  of chlorophyll a/b light-harvesting complex in thylakoids and aggregates:
  information from circular differential scattering,'' {\em Biochemistry},
  vol.~27, no.~7, pp.~2425--2430, 1988.

\bibitem{Finzi1989}
L.~Finzi, C.~Bustamante, G.~Garab, and C.-B. Juang, ``Direct observation of
  large chiral domains in chloroplast thylakoid membranes by differential
  polarization microscopy,'' {\em Proceedings of the National Academy of
  Sciences}, vol.~86, no.~22, pp.~8748--8752, 1989.

\bibitem{Dobrikova2003}
A.~G. Dobrikova, Z.~V{\'a}rkonyi, S.~B. Krumova, L.~Kov{\'a}cs, G.~K. Kostov,
  S.~J. Todinova, M.~C. Busheva, S.~G. Taneva, and G.~Garab, ``Structural
  rearrangements in chloroplast thylakoid membranes revealed by differential
  scanning calorimetry and circular dichroism spectroscopy. thermo-optic
  effect{\textdagger},'' {\em Biochemistry}, vol.~42, pp.~11272--11280, sep
  2003.

\bibitem{Jajoo2012}
A.~Jajoo, M.~Szab{\'{o}}, O.~Zsiros, and G.~Garab, ``Low {pH} induced
  structural reorganization in thylakoid membranes,'' {\em Biochimica et
  Biophysica Acta ({BBA}) - Bioenergetics}, vol.~1817, no.~8, pp.~1388--1391,
  2012.

\bibitem{Staehelin1986}
L.~Staehelin, ``Chloroplast structure and supramolecular organization of
  photosynthetic membranes,'' in {\em Photosynthesis III}, pp.~1--84, Springer,
  1986.

\bibitem{Palmer1986}
J.~D. Palmer and D.~B. Stein, ``Conservation of chloroplast genome structure
  among vascular plants,'' {\em Current Genetics}, vol.~10, no.~11,
  pp.~823--833, 1986.

\bibitem{Minagawa2013}
J.~Minagawa, ``Dynamic reorganization of photosynthetic supercomplexes during
  environmental acclimation of photosynthesis,'' {\em Frontiers in Plant
  Science}, vol.~4, p.~513, 2013.

\bibitem{Nagy2014}
G.~Nagy, R.~{\"U}nnep, O.~Zsiros, R.~Tokutsu, K.~Takizawa, L.~Porcar, L.~Moyet,
  D.~Petroutsos, G.~Garab, G.~Finazzi, {\em et~al.}, ``Chloroplast remodeling
  during state transitions {inChlamydomonas} reinhardtiias revealed by
  noninvasive techniques in vivo,'' {\em Proceedings of the National Academy of
  Sciences}, vol.~111, pp.~5042--5047, mar 2014.

\bibitem{Murakami1970}
S.~Murakami and L.~Packer, ``Light-induced changes in the conformation and
  configuration of the thylakoid membrane of ulva and porphyra chloroplasts in
  vivo,'' {\em Plant Physiology}, vol.~45, no.~3, pp.~289--299, 1970.

\bibitem{Schlodder2007}
E.~Schlodder, V.~V. Shubin, E.~El-Mohsnawy, M.~Roegner, and N.~V. Karapetyan,
  ``Steady-state and transient polarized absorption spectroscopy of photosytem
  i complexes from the cyanobacteria arthrospira platensis and
  thermosynechococcus elongatus,'' {\em Biochimica et Biophysica Acta ({BBA}) -
  Bioenergetics}, vol.~1767, no.~6, pp.~732--741, 2007.

\bibitem{Bekasova2013}
O.~Bekasova, V.~Shubin, I.~Safenkova, L.~Kovalyov, and B.~Kurganov,
  ``Structural changes in {R-phycoerythrin} upon {CdS} quantum dot synthesis in
  tunnel cavities of protein molecules,'' {\em International Journal of
  Biological Macromolecules}, vol.~62, pp.~623--628, 2013.

\bibitem{Berkaloff1983}
C.~Berkaloff, J.~Duval, N.~Hauswirth, and B.~Rousseau, ``Freeze fracture study
  of thylakoids of {Fucus} serratus,'' {\em Journal of Phycology}, vol.~19,
  no.~1, pp.~96--100, 1983.

\bibitem{DeMartino2000}
A.~De~Martino, D.~Douady, M.~Quinet-Szely, B.~Rousseau, F.~Cr{\'e}pineau,
  K.~Apt, and L.~Caron, ``The light-harvesting antenna of brown algae,'' {\em
  The FEBS Journal}, vol.~267, no.~17, pp.~5540--5549, 2000.

\bibitem{Buechel1997}
C.~B{\"u}chel and G.~Garab, ``Organization of the pigment molecules in the
  chlorophyll a/c light-harvesting complex of {Pleurochloris} meiringensis
  ({Xanthophyceae}). characterization with circular dichroism and absorbance
  spectroscopy,'' {\em Journal of Photochemistry and Photobiology B: Biology},
  vol.~37, no.~1-2, pp.~118--124, 1997.

\bibitem{Szabo2008}
M.~Szab{\'o}, B.~Lepetit, R.~Goss, C.~Wilhelm, L.~Must{\'a}rdy, and G.~Garab,
  ``Structurally flexible macro-organization of the pigment--protein complexes
  of the diatom {Phaeodactylum} tricornutum,'' {\em Photosynthesis Research},
  vol.~95, no.~2-3, pp.~237--245, 2008.

\bibitem{Nagy2012}
G.~Nagy, M.~Szab{\'o}, R.~{\"U}nnep, G.~K{\'a}li, Y.~Miloslavina, P.~H.
  Lambrev, O.~Zsiros, L.~Porcar, P.~Timmins, L.~Rosta, {\em et~al.},
  ``Modulation of the multilamellar membrane organization and of the chiral
  macrodomains in the diatom phaeodactylum tricornutum revealed by small-angle
  neutron scattering and circular dichroism spectroscopy,'' {\em Photosynthesis
  research}, vol.~111, no.~1-2, pp.~71--79, 2012.

\bibitem{Lepetit2007}
B.~Lepetit, D.~Volke, M.~Szab{\'o}, R.~Hoffmann, G.~Garab, C.~Wilhelm, and
  R.~Goss, ``Spectroscopic and molecular characterization of the oligomeric
  antenna of the diatom {Phaeodactylum} tricornutum,'' {\em Biochemistry},
  vol.~46, no.~34, pp.~9813--9822, 2007.

\bibitem{Buechel2003}
C.~B{\"u}chel, ``Fucoxanthin-chlorophyll proteins in diatoms: 18 and 19 {kDa}
  subunits assemble into different oligomeric states,'' {\em Biochemistry},
  vol.~42, no.~44, pp.~13027--13034, 2003.

\bibitem{Katoh1989}
T.~Katoh, M.~Mimuro, and S.~Takaichi, ``Light-harvesting particles isolated
  from a brown alga, {Dictyota} dichotoma: A supramolecular assembly of
  fucoxanthin-chlorophyll-protein complexes,'' {\em Biochimica et Biophysica
  Acta (BBA)-Bioenergetics}, vol.~976, no.~2-3, pp.~233--240, 1989.

\bibitem{Ghazaryan2016}
A.~Ghazaryan, P.~Akhtar, G.~Garab, P.~H. Lambrev, and C.~B{\"u}chel,
  ``Involvement of the {Lhcx} protein {Fcp6} of the diatom {Cyclotella}
  meneghiniana in the macro-organisation and structural flexibility of
  thylakoid membranes,'' {\em Biochimica et Biophysica Acta
  (BBA)-Bioenergetics}, vol.~1857, no.~9, pp.~1373--1379, 2016.

\bibitem{Steinbach2014}
G.~Steinbach, K.~Pawlak, I.~Pomozi, E.~A. T{\'{o}}th, A.~Moln{\'{a}}r,
  J.~Matk{\'{o}}, and G.~Garab, ``Mapping microscopic order in plant and
  mammalian cells and tissues: novel differential polarization attachment for
  new generation confocal microscopes ({DP}-{LSM}),'' {\em Methods and
  Applications in Fluorescence}, vol.~2, no.~1, p.~015005, 2014.

\bibitem{Gombos2008}
I.~Gombos, G.~Steinbach, I.~Pomozi, A.~Balogh, G.~V{\'a}mosi, A.~Gansen,
  G.~L{\'a}szl{\'o}, G.~Garab, and J.~Matk{\'o}, ``Some new faces of membrane
  microdomains: a complex confocal fluorescence, differential polarization, and
  fcs imaging study on live immune cells,'' {\em Cytometry Part A}, vol.~73,
  no.~3, pp.~220--229, 2008.

\bibitem{Dierssen2015}
H.~Dierssen, A.~Chlus, and B.~Russell, ``Hyperspectral discrimination of
  floating mats of seagrass wrack and the macroalgae {Sargassum} in coastal
  waters of {Greater Florida Bay} using airborne remote sensing,'' {\em Remote
  Sensing of Environment}, vol.~167, pp.~247--258, 2015.

\end{thebibliography}

\bibliographystyle{ieeetr}

\end{document}